# Long Distance, Unconditional Teleportation of Atomic States Via Complete Bell State Measurements

S. Lloyd and M.S. Shahriar
Research Laboratory of Electronics and Dept. of Mechanical Engineering,
Massachusetts Institute of Technology,
Cambridge, MA 02139

P.R. Hemmer
Air Force Research Laboratory, Sensors Directorate,
Hanscom Air Force Base, MA 01731

## Abstract

This paper proposes a scheme for creating and storing quantum entanglement over long distances. Optical cavities that store this long-distance entanglement in atoms could then function as nodes of a quantum network, in which quantum information is teleported from cavity to cavity. The teleportation can be carried out unconditionally via measurements of all four Bell states, using a method of sequential elimination.

Quantum computers communicating with each other reliably over macroscopic distances would be able to perform both quantum and classical computations in a secure fashion [1]. This paper proposes a robust scheme for constructing such a quantum network. A method is developed for creating entanglement between two distant atoms using entangled photons. Any two atoms that share entanglement can exchange quantum information by the process of teleportation [2]. Quantum information processing need only be performed locally, within the nodes. The scheme should allow reliable transmission of quantum information between quantum microcomputers separated by distances of hundreds of kilometers.

In general, it is difficult to create a quantum wire that can transmit quantum information [3]. Direct quantum communication is fragile; although methods exist for coping with noisy quantum channels, they are complicated and require considerable computational time [4]. The solution is to create a quantum network that does not require reliable quantum wires [5]. Cavity quantum electrodynamics provides mechanisms for communicating between cavities using inter-cavity photons [1,5]. The key technology proposed here is a method for transmitting quantum entanglement over long distances, capturing it in optical cavities, and storing it in atoms.

We describe the method in general terms and provide specific details later. First, use parametric down conversion to create pairs of photons that are entangled both in momentum and in polarization. Send one photon to cavity 1 and the other to cavity 2, equidistant from the source. Each cavity contains an ultra-cold atom, trapped in an optical potential. Because of their momentum entanglement, each of the entangled pair of photons arrives at its respective cavity at the same time. Although many (indeed most) of the photon pairs will fail to arrive and enter their respective cavities, on occasion a photon will enter cavity 1 and its entangled pair photon will enter cavity 2 at the same time. Once in the cavity, the photon can drive a transition between the A and the degenerate B levels of the atom (see figure 1a). This effectively transfers the photon entanglement to the degenerate B levels of the atoms in cavities 1 and 2.

This entanglement can be detected and stored as follows. Concentrate first on a single cavity. To protect the quantum information in the event that the atom has absorbed the photon, drive a coherent transition from the degenerate B levels to the long-lived D hyperfine levels as in figure 1a. Now detect whether or not the atom has absorbed a photon by driving a cycling transition from the A to the C level of the atom. If no fluorescence is seen, then the atom successfully absorbed the photon and the resulting entanglement in D will be stored for subsequent manipulations. On the other hand, if fluorescence is observed, then the atom was still in A, which means it failed to absorb the photon. In this case the atom will be allowed to return to A (from C), ready to absorb the next photon that enters the cavity. When the keeper of cavity 1 has successfully captured a photon, she notes the time of arrival and calls the keeper of cavity 2 over the phone to see whether he has successfully captured a photon at the same time. If not, she returns her atom to the state A and tries again. If both cavity keepers have captured a photon at the same time, however, they now possess two distantly separated atoms that are entangled.

The above method allows the creation of entanglement between atoms separated by long distances in the face of arbitrarily high loss rates, in principle. In practice, the fidelity of the entanglement created is determined by the accuracy with which one is able to perform the various operations. First, account for the losses with probability $\lambda$ that a photon survives its trek from source to cavity and is absorbed by an atom. We can write $\lambda = 10^{-L/10}$, where $L$ is the net loss in decibels. If $\lambda$ is small, most of the time the atom will be in A, and the fluorescence will be observed on the cycling transition. Occasionally however, the detector will fail to detect this fluorescence and erroneously conclude that the atom is in D. The false detection probability can be written as $\varepsilon = p^N$ where $p$ is the probability that the detector fails to detect a single photon and $N$ is the number of times the cycling transition is repeated. Alternatively, the false-positive probability can be expressed in terms of the (effective) signal-to-noise ratio of the cycling transition $S$ as $\varepsilon = 10^{-S/10}$. The fidelity is high as long as $\varepsilon << \lambda$, which means that when the cavity keepers believe that they have both captured photons, they are almost always right. Clearly the fidelity can be increased to the desired level by driving the cycling transition longer. Of course, this is limited by the probability of driving an off-resonant transition to a non-cycling state. Fortunately, in practice not many cycles are needed. The main problem with high losses is that the entanglement capture procedure must be repeated $\sim 1/\lambda^2 =$

$10^{L/5}$ times in order to capture a single entangled pair. This will be discussed in more detail below.

Fidelity can also be degraded by spontaneous emission from the excited state, but this can be minimized by using optically off resonance (OOR) Raman excitation. To improve entanglement fidelity further, many atoms can be included in the trap and entanglement purification techniques can be used [6]. Finally, to store and maintain the entanglement for times long compared with the decoherence time of the atoms, one can in principle combine entanglement purification together with the use of quantum error correcting codes [7]. However, because of the long coherence time of atoms in the proposed method this should not be needed.

Let us now look at how such a scheme might be carried out in practice using rubidium atoms (figure 1b). A UV laser will be used to excite a non-linear crystal. Via type-II parametric down-conversion, this crystal will produce pairs of entangled photons, each at 795 nm. A bright, narrow-band source of photon pairs is described in ref. 8. We consider the polarization entanglement to be of the form $(|\sigma_+\rangle_1|\sigma_-\rangle_2+|\sigma_-\rangle_1|\sigma_+\rangle_2)/\sqrt{2}$, where $\sigma_+(\sigma_-)$ indicates right(left) circular polarization. The protocol will work for any arbitrary but known phase between the two components as well. Each beam is coupled into a fiber, and transported to an optical cavity with slow decay and a strong vacuum rabi frequency (20 MHz) [9].

Each cavity, at its center, holds a single rubidium atom, confined by a focused $CO_2$ laser. The mean number of atoms caught is controlled via the parameters involved in the process. Starting with a mean-value of about 5, the atoms will be knocked out one by one while monitoring the remaining fluorescence until only a single atom remains. It has been demonstrated recently [10] that at a pressure of $10^{-11}$ Torr, atoms survive for more than 2 minutes in a $CO_2$ trap. The trap lifetime, and hence the decoherence time, can be increased up to an hour by housing the trap chamber in a liquid helium cryostat.

To load the photon into the cavity, one approach, akin to ref. 1, is to use a carefully timed and shaped Raman pulse to induce the simultaneous admission of the photon into the cavity and absorption by the atom. In the more general case, when the temporal envelope of the photon cannot be controlled, the cavity will still load with some probability $\eta$, which can be enhanced by proper choice of cavity and laser pulse (B-D transition) parameters. It is important to note that due to the momentum entanglement the probability of absorbing two conjugate photons is also $\eta$, not $\eta^2$[11].

Figure 2 illustrates the energy levels and transitions to be employed. Initially, the atom(s) are prepared to be in the F=1, $m_F$=0 ground state (`A' level). The photon excites the dashed transitions to the F=1, $m_F$=±1 excited level (`B' levels) (fig 2a). A $\pi$ polarized beam completes the Raman excitation, putting the atom in a superposition of the F=2, $m_F$=±1 ground states ('D' levels). Note that because of the large detuning of the $CO_2$ laser, atoms in both the 'A' levels and the 'D' levels will remain trapped. To detect whether the photon has been absorbed by the atom, the F=1 ground state is detected by exciting the cycling transition ('A to C') shown in figure 2b. The same process occurs at the other cavity, and the results sent to a co-incidence monitor.

Let us now insert reasonable numbers for losses and errors and see how far this method is likely to operate without the use of quantum repeaters [12]. Realistic optical fiber losses for 795 nm photons are 3dB per kilometer. Putting the two cavities 10 kilometers apart gives a per-cavity loss rate of $L \sim 15$ dB, assuming that losses are dominated by the fiber. Sensitive photodetectors along with appropriate optics can give a photon detection rate of 25%. Hence, driving the cycling transition 30 times gives a signal to noise of ~ 37 dB, more than adequate to compensate for fiber loss. The $CO_2$ trap can be made deep enough to overcome the heating expected from this process. The entanglement creation rate can be estimated by multiplying the fluorescent photon emission time (30 nanoseconds) by the number of cycles and the mean number of times the capture procedure must be repeated ($10^{2L/10}$) to yield one pair per millisecond. Accordingly, a distance of ten kilometers between nodes of the quantum network seems a reasonable objective.

The two cavities now contain atoms that are entangled with respect to long-lived states. This entanglement can be used for quantum teleportation, quantum cryptography, superdense coding, or remote clock synchronization[13], for example. Here, we show an explicit construction for performing the teleportation of a quantum state (fig. 3). The entangled atoms are atom 2 (with Alice) and atom 3 (with Bob). Alice has a third atom (atom 1), whose quantum state she will teleport. Atom 1 is trapped in another node of the $CO_2$ beam, and shares a second cavity with atom 2 (figure 3a). We adopt the abbreviated state designations for the $5^2S_{1/2}$ sublevels: $|a\rangle \equiv |F=2, m_F=-1\rangle$, $|b\rangle \equiv |F=2, m_F=+1\rangle$, $|c\rangle \equiv |F=1, m_F=-1\rangle$, and $|d\rangle \equiv |F=1, m_F=+1\rangle$.

The state of atoms 2 and 3 can be written as $|\psi_{23}\rangle = \{|a\rangle_2|b\rangle_3 + |b\rangle_2|a\rangle_3\}/\sqrt{2}$, assuming that there is no birefringence during the propagation of the photons. Alice can put atom 1 into an unknown state $|\varphi_1\rangle = \{\alpha|c\rangle_1 + \beta|a\rangle_1\}$ by using an OOR Raman pulse of unmeasured duration that couples $|c\rangle_1$ to $|a\rangle_1$. Explicitly, $\alpha(t)=\alpha_0$ and $\beta(t)=\beta_0\theta(t)$, where $\theta(t)=\exp[-i(\omega_M t+\xi)]$, the frequency $\omega_M$ and phase $\xi$ being determined by those of the oscillator used to generate the second Raman frequency from the first. Using the scheme of Pellizzari et al. [14], which can be realized using the transitions shown in figure 3b, Alice transfers the state of atom 1 into atom 2, leaving atom 1 in a pure state $|c\rangle_1$, and atoms 2 and 3 in the state $|\phi_{23}\rangle = \{|A_+\rangle(\alpha_0|b_3\rangle+\beta_0|a_3\rangle) + |A_-\rangle(\alpha_0|b_3\rangle-\beta_0|a_3\rangle) + |B_+\rangle(\beta_0|b_3\rangle+\alpha_0|a_3\rangle) + |B_-\rangle(-\beta_0|b_3\rangle+\alpha_0|a_3\rangle)\}/2$, where the Bell basis states are given by $|A_\pm\rangle = \{|c_2\rangle \pm \theta|b_2\rangle\}/\sqrt{2}$, and $|B_\pm\rangle = \{|d_2\rangle \pm \theta|a_2\rangle\}/\sqrt{2}$.

To measure the Bell states, Alice first applies a set of pulses that maps the Bell states to bare atomic states. Consider the states $|A_\pm\rangle$. Alice applies an OOR Raman $\pi/2$ pulse, coupling $|c_2\rangle$ to $|b_2\rangle$, using a $\sigma_+$ polarized light at $\omega_1$ and a $\sigma_-$ polarized light at $\omega_2$, where $\omega_1 - \omega_2 = \omega_M$ (fig. 4a). The off resonant pulses are tuned near the F=1 excited state to avoid interactions between $|a_2\rangle$ and $|d\rangle_2$. She generates $\omega_2$ from $\omega_1$ using an oscillator with a known phase shift of $\xi$. She chooses $\xi=-\pi/2$, converting $|A_+\rangle$ to $|c_2\rangle$, and $|A_-\rangle$ to $|b_2\rangle$. Similarly, she applies another OOR Raman $\pi/2$ pulse with different polarizations to convert $|B_+\rangle$ to $|d_2\rangle$, and $|B_-\rangle$ to $|a_2\rangle$.

Alice can now measure the internal states by using a method of *sequential elimination*. First, she applies a Raman pulse to transfer the amplitude of state $|d_2\rangle$ to an auxiliary state in the $5^2S_{1/2}$, F=2 manifold (fig. 4b). A magnetic field can be applied in order to provide the necessary spectral selectivity. She then probes the amplitude of state $|c_2\rangle$ by driving the $5^2S_{1/2}$, F=1 to $5^2P_{3/2}$, F=0 cycling transition. If she detects fluorescence, she concludes that the atom is in state $|c_2\rangle$, which in turn means she has measured the Bell state $|A_+\rangle$. If she fails to see fluorescence, then she has eliminated this possibility, and now applies a Raman pulse to return the amplitude of the auxiliary state to state $|d_2\rangle$. She again drives the cycling transition, and detection of fluorescence implies she has measured the Bell state $|B_+\rangle$. Otherwise, she applies a set of Raman pulses to transfer the amplitude of $|a_2\rangle$ to $|c_2\rangle$ and $|b_2\rangle$ to $|d_2\rangle$. She now repeats the detection scheme for state $|c_2\rangle$. If she sees fluorescence, the atom is in $|c_2\rangle$, which implies that she has measured $|A_-\rangle$. If not, she has eliminated three possibilities, which means that the system is in $|B_-\rangle$.

Alice now sends a two-bit classical message to Bob, informing of which state she has found the world to be in. Bob now has to make some transformations to his atom in order to produce the state $|\phi_1\rangle$ in atom 3. This he can accomplish as follows. If Alice found $|A_+\rangle$, Bob does nothing, and atom three is already in state $|\phi_1\rangle$. On the other hand, if Alice found $|A_-\rangle$, then Bob has to flip the phase of $|a_3\rangle$ by $\pi$ with respect to $|b_3\rangle$. This can be achieved by applying an OOR Raman $2\pi$ pulse connecting $|a_3\rangle$ to the auxiliary ground state $|F=1, m_F=0\rangle$ via the $|F=0, m_F=0\rangle$ state in $5^2P_{3/2}$. Atom 3 is now in state $|\phi_1\rangle$, as desired. If Alice found $|B_+\rangle$, then Bob first applies an OOR Raman $\pi$ pulse coupling $|a_3\rangle$ to $|b_3\rangle$ to swap their amplitudes, again producing the desired state. Finally, if Alice found $|B_-\rangle$, then Bob first applies a $\pi$ pulse as above to swap amplitudes, followed by the $2\pi$ pulse for the $\pi$ phase change, producing the desired state.

Finally, we point out that the teleportation protocol presented here can also be implemented with only two atoms in a cavity, and without the entangled photon pairs. First, the necessary entanglement between the two atoms can be realized using the scheme of ref. 14. Second, the quantum state to be teleported can be encoded directly into one of the entanglement atoms, using the additional sublevels.

This work was supported by AFOSR grant # F49620-98-1-0313 and ARO grant #s DAAG55-98-1-0375 and DAAD19-001-0177

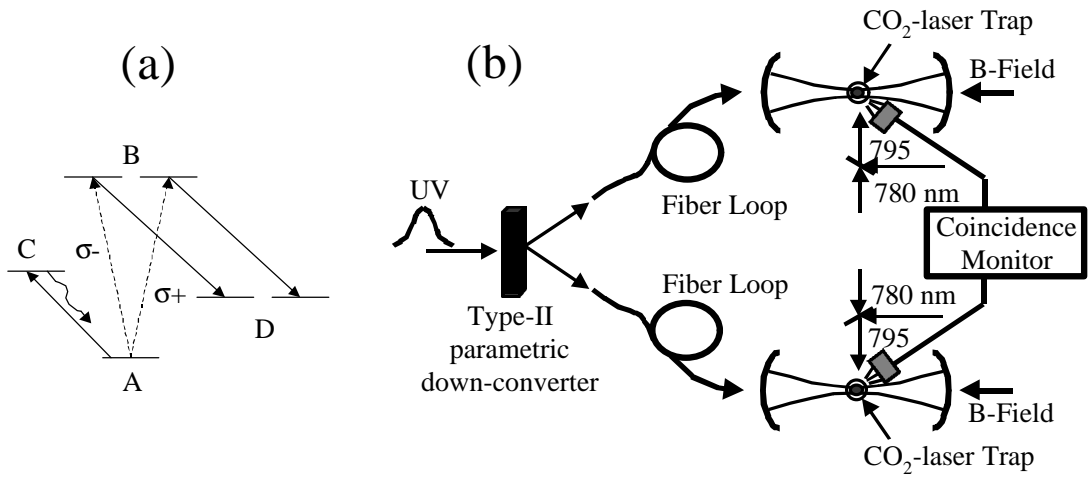

Figure 1. Schematic illustration of the proposed experiment for creating potentially long distance entanglement between a pair of trapped rubidium atoms (see text)

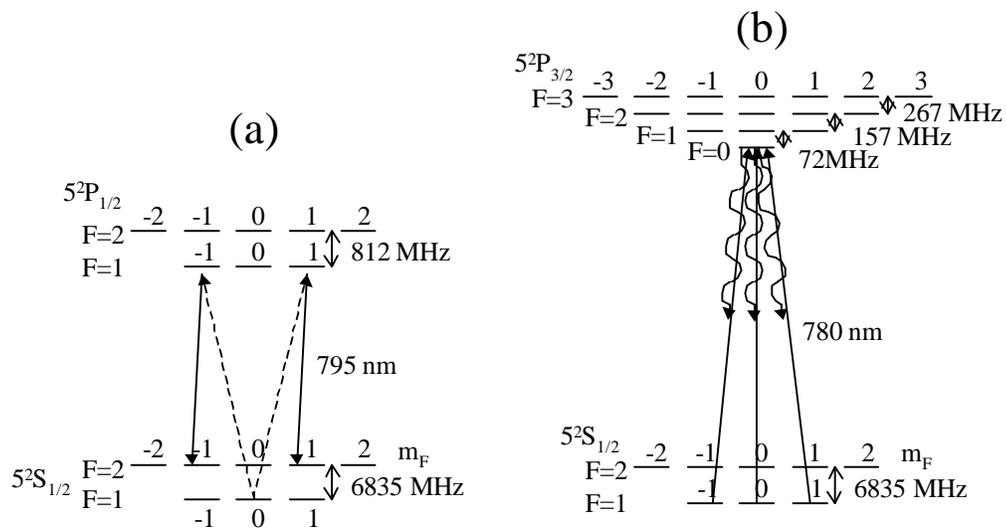

Figure 2. Schematic illustration of the steps to be used in storing quantum coherence in a rubidium atom, and detecting it non-destructively (see text).

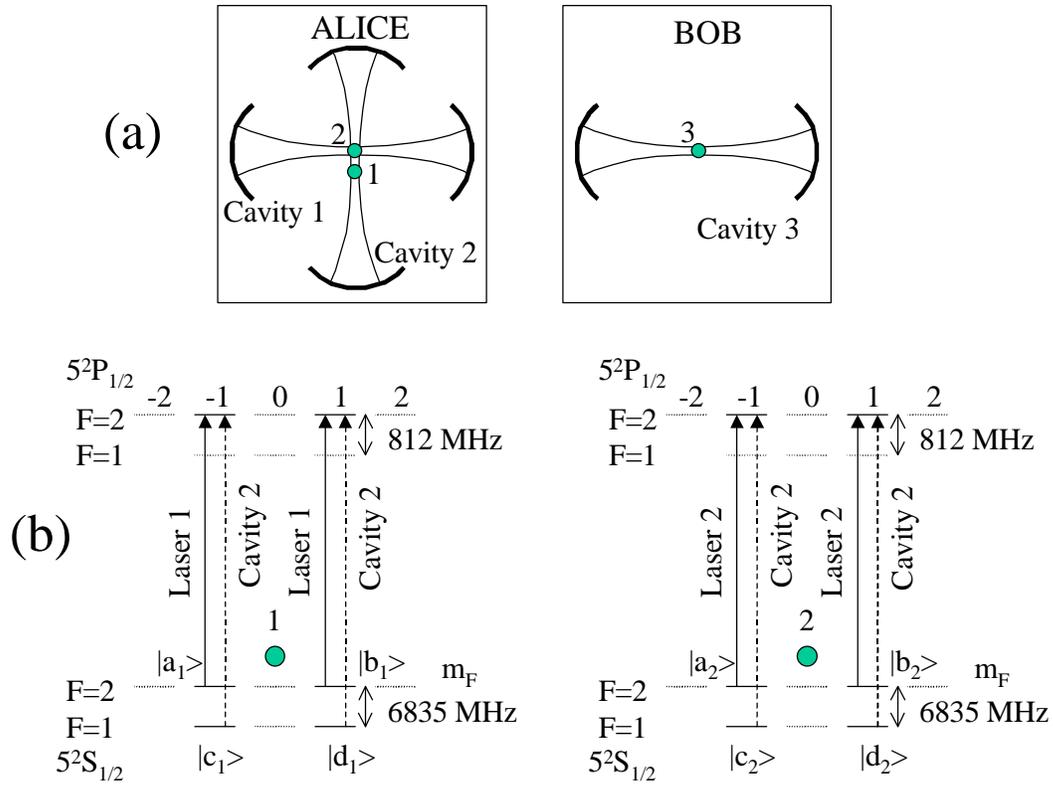

Figure 3: Teleportation using trapped Rb atoms: (a) Atom 2 (with Alice) and atom 3 (with Bob) are entangled via storage of the entangled photon pairs, as described above. Atom 1 (also with Alice) is held by a second $CO_2$ laser node, and shares a common cavity that is orthogonal to the one used for capturing the entangled photons. (b) Basic model for Alice to transfer the coherence from atom 1 to atom 2, in preparation for teleportation.

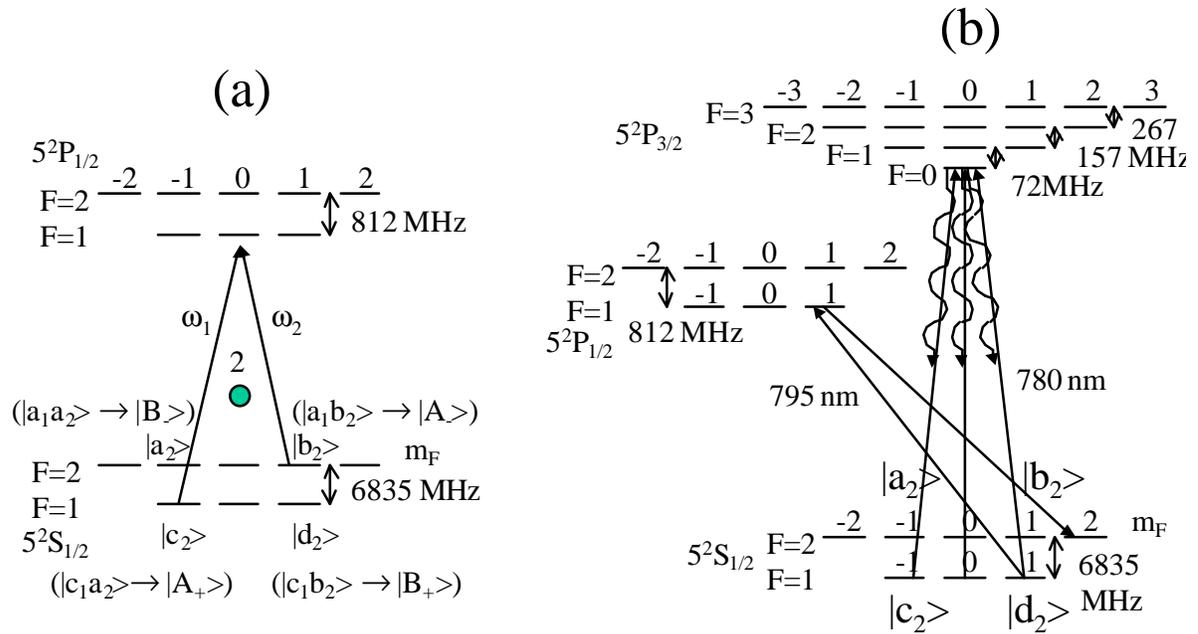

Figure 4. Bell state measurement: (a) Off resonant Raman transition used to map the amplitudes of the Bell states onto the bare atomic states (b) Bell state detection is done sequentially using Raman transitions (see text).